# Specification-first convergence with an AI coding agent: a case study of dismantling a core architectural invariant across 189 files in a 717k-line codebase with no test oracle and no human code review

Joël Abenhaïm
*AI Sovereign Labs, Paris, France*
joel@aisovereignlabs.ai
31 July 2026

## Abstract

This paper reports a single, fully instrumented case study of a large-scale architectural refactoring by an AI coding agent under a specification-first protocol, with no human review of the generated code and no pre-existing oracle to validate the target behaviour. The task, dismantling a central invariant across a large interdependent codebase, was assessed by the author as effectively infeasible through incremental refactoring, the kind of change that conventionally calls for a rewrite instead. Under the protocol described here, the agent completed it successfully.

The system is a 717,725-line production TypeScript application across 3,648 files. The task required dismantling a core lifetime invariant: the guarantee that a UI panel remains open for the duration of an AI request. The target behaviour was that a streaming generation survives the closing of its panel and can be reattached, on reopening, to the same live stream with no loss or duplication.

The protocol: formal specification by the agent, 14 refinement cycles auditing that specification against the source code, atomic implementation, a compile/test feedback loop, then 17 verification cycles auditing the code against the frozen specification. Across 31 audit passes, 201 defects were corrected before any human executed the program. The convergence criterion was empirical: two consecutive verification passes returning zero findings.

The change touched 189 files (31 new); with the extraction phase, the two commits total 288 files, 34,770 insertions, 16,422 deletions. Across the first and roughly thirty later sessions, the software behaved as specified, no bug observed. Elapsed: three days; cost: USD 2,430.

The full specification and raw session logs, 1,500+ pages in French, are published as evidence, allowing inspection of the process and submission to a language model for consistency checking.

## 1. Introduction

Current AI coding agents, including Claude Code, Codex, Copilot and Cursor, demonstrate high throughput on isolated, self-contained tasks. Human review of the generated code is considered necessary in most practices.

Review has an inherent scaling limit. When a single change spans hundreds of interdependent files and must preserve dozens of implicit invariants simultaneously, the review itself becomes the bottleneck, and beyond a certain size it stops being a realistic quality gate at all: no reviewer holds the entire dependency graph of such a change in working memory. A recent systematic review formalises this as a review bottleneck: telemetry across thousands of developers shows AI-assisted work producing more pull requests but review times increasing by up to 91%, with flat delivery metrics overall [1].

This paper reports evidence that control can be relocated instead. Rather than auditing the generated code after the fact, the intent is audited before generation, in the form of a written specification that is repeatedly re-checked against the real source code until it stops producing findings. Code is then generated against a frozen reference, and audited again against that same reference.

This paper documents one execution of that protocol, chosen specifically for its difficulty as a stress test of the method rather than as a representative sample of routine maintenance work: the removal of a lifetime invariant in a live streaming subsystem, a class of change that ordinarily invites a rewrite rather than a refactor, completed here in three days.

## 2. Related work

The dominant evaluation paradigm for coding agents is exemplified by SWE-bench [2], which measures an agent's ability to resolve real-world GitHub issues by producing a patch that passes a held-out, human-written test suite. This formulation assumes an oracle: the correct behaviour is already encoded in existing tests, and the agent's task is to find code that satisfies them. SWE-bench Pro [3] extends this paradigm to longer-horizon, more industrially representative tasks, and reports that even under a unified scaffold, current systems resolve fewer than 45% of instances.

The task examined in this paper does not fit that formulation: no pre-existing test suite encoded the target behaviour, because that behaviour, a generation surviving the closing of its panel, did not exist before the change. There is no oracle to satisfy, only a specification to construct and then hold the implementation to.

Separately, SWE-agent [4] established that the interface through which an agent reads and edits a codebase materially affects its reliability, independent of the underlying model, a finding consistent with the execution model reported in Section 5, in which the agent proposes patches rather than writing files directly. OpenAI's harness engineering report [5] documents a comparable shift in review practice, without addressing the specification-refinement stage examined here.

More recently, industrial practice converged on a pattern named loop engineering, coined by Addy Osmani in June 2026 [6] building on public statements by Peter Steinberger and Anthropic's Boris Cherny. Instead of prompting an agent turn by turn, the developer builds an outer system that finds the work, hands it to the agent, checks the result with a second agent, and records what is done outside the model's context.

Anthropic published its own documentation of the practice the same month [7], describing four loop types and recommending a second agent with no memory of the change for review.

That pattern needs a way to tell automatically whether one unit of work is finished. The example given for its most autonomous form is a passing test suite and a clean linter [6]. It works when the work splits into units that can each be checked on their own. It says nothing about how to get one large change right when the units cannot be separated and no test can tell a correct implementation from an incorrect one. That is the case examined in this paper. The two approaches address different problems rather than competing.

The pattern also holds that the agent writing the code must not be the agent checking it: Anthropic's guidance recommends a second agent with no memory of the change, on the grounds that a reviewer with fresh context is less biased and not influenced by the first agent's reasoning [7]. The protocol described in Section 5 applies that principle differently. The checker is a fresh session of the same agent, not a different agent with different instructions, comparing the code to a document written and frozen before the code existed. The separation this protocol relies on is in that referent, not in the identity of the reviewer. This is consistent with the finding that a model revising its own output with nothing external to compare against does not improve, and can get worse [8].

Beyond the identity of the checker, the sheer volume of the harness also matters. The agent used in this study operates under a system prompt of approximately 250,000 characters, encoding hundreds of rules accumulated from observed failures during construction and corrected incrementally to reduce the need for manual oversight. Davis et al. [9] report a comparable pattern at a different scale: their governance substrate, spanning agent documentation, static and dynamic analyses, and tooling, reached 1.16 MLOC, accumulated across a 12-week project as recurring agent failures were converted into durable mechanisms. Both approaches converged empirically toward the same direction: lengthening the constraints given to the agent, reaching substantial proportions in both cases, in order to automate more of the work and reduce the costly need for manual review. Their catalogue of mechanisms is documented in full on a companion site [10].

Industrial deployments of AI coding agents at scale have also been documented, addressing a complementary problem. Cloudflare's AI-orchestrated code review system processes over 48,000 merge requests across 5,169 repositories using seven domain-specialised reviewer agents, built as an assistive layer over human-authored code rather than a code-generation system [11]. Anthropic's 2026 agentic coding report documents production deployments including a seven-hour single-run implementation on the 12.5M-line vLLM codebase, validated against a pre-existing numerical accuracy reference [12]. Both represent mature, high-volume industrial use of AI in the software lifecycle, in review and in generation respectively.

The strongest industrial demonstration to date is the Bun rewrite: 535,496 lines of Zig ported to Rust in 11 days by up to 64 Claude agents in parallel, over a million generated lines passing the full pre-existing test suite and shipping to production [13]. The verification stack that made this mergeable is anchored on oracles: a million-assertion test suite independent of the implementation language, the Rust compiler, and adversarial agent review of every change. The task examined here has no such oracle: the target behaviour did not exist before the change, so the referent had to be constructed instead.

This paper's contribution sits alongside, on a specific and narrower configuration: an agent generating code across hundreds of interdependent files with no human review of the output at any stage, on a task with no pre-existing oracle to check correctness against.

## 3. System under study

The codebase is a VS Code extension implementing an AI coding agent [14], of the same functional category and comparable size to commercially available agents. At the start of the operation it comprised 717,725 lines of TypeScript across 3,648 files. It was written and is maintained by a single developer, who is also the author of this paper, under the same specification-first methodology, across roughly 4,000 commits; this paper instruments a single one of those operations in full.

Three components are relevant to the task:

- An orchestration backend, running in the extension host process, responsible for issuing model requests and streaming their output.
- A React frontend, running in a webview, responsible for rendering the stream: response text, reasoning traces, tool calls, code blocks and other interface elements interleaved in a single composite view.
- A custom RPC data bus connecting the two processes.

The system is proprietary and closed-source, custom end to end. Its code was never published, which excludes the possibility that the agent reproduced a solution present in its training data.

## 4. Task definition

The request was submitted to the agent as follows, in a single natural-language prompt with no accompanying formal specification:

> *"Analyze this request in detail. Produce a comprehensive report covering understanding and feasibility.*
>
> *- The system that streams an active session on the frontend within EasyAgents: What is it? A complex system that distinguishes between current and past messages; it is tricky to manage, involves handling streaming pauses, includes mechanisms to prevent data loss due to race conditions, etc. It requires detailed study, it is extremely complex and full of pitfalls, not something to be looked at casually. We will start by analyzing it closely, then extract the code and create technical documentation for it within a new dedicated module (likely located somewhere under `orch/`). "Extract" means moving the code while making it generic, ready for use in both the EasyAgents use case and the standard session use case. The extraction must be done properly, including all associated systems, no sloppy coding. A successful extraction at this stage means the code is separated and reconnected to EasyAgents without EasyAgents actually containing the code itself; the visible functionality remains the same (the change is invisible on screen but real in the codebase).*
>
> *- What needs to be done? Extract it from EasyAgents, move it, and make it generic.*
>
> *- What is the goal? It must work with standard AI-code sessions.*
>
> *- What is the result? Closing a standard AI-code session view no longer kills the orchestrator process; the session remains visible in the activity sidebar.*

*- Why? We sometimes encounter a bug that turns the view gray (an Electron memory overflow issue in VS Code), which is annoying because nothing is visible. Currently, we have to let it finish blindly until the activity view indicates completion, or kill the request, but that means losing the work done and having to restart the last user message from scratch. With this new system, if the bug occurs, we can close the view and reopen it by clicking on it in the activity list. Reopening it restores the display.*

*- UX adaptation required: Since the process isn't killed when the view closes, we need a way to kill it. In the activity view, each item currently has two lines (title and elapsed time). Add a third line containing a stop button (stop icon, small size, danger style). This button should appear only when the view is hidden but the process is still running; it (and the third line) should not appear if the process finishes normally or if the view is displayed again. The button becomes disabled after being clicked to prevent accidental double-clicks and bugs. Clicking it kills the process, just as closing the window currently does."*

The functional requirement can be restated as follows. Closing the panel of a running session must no longer terminate the underlying generation. The session must remain visible in an activity sidebar, with a detached stop control appearing only while the panel is closed and the source still running. Reopening the panel must restore the display to a state indistinguishable from one in which the panel had never been closed, then resume live streaming without loss or duplication of tokens.

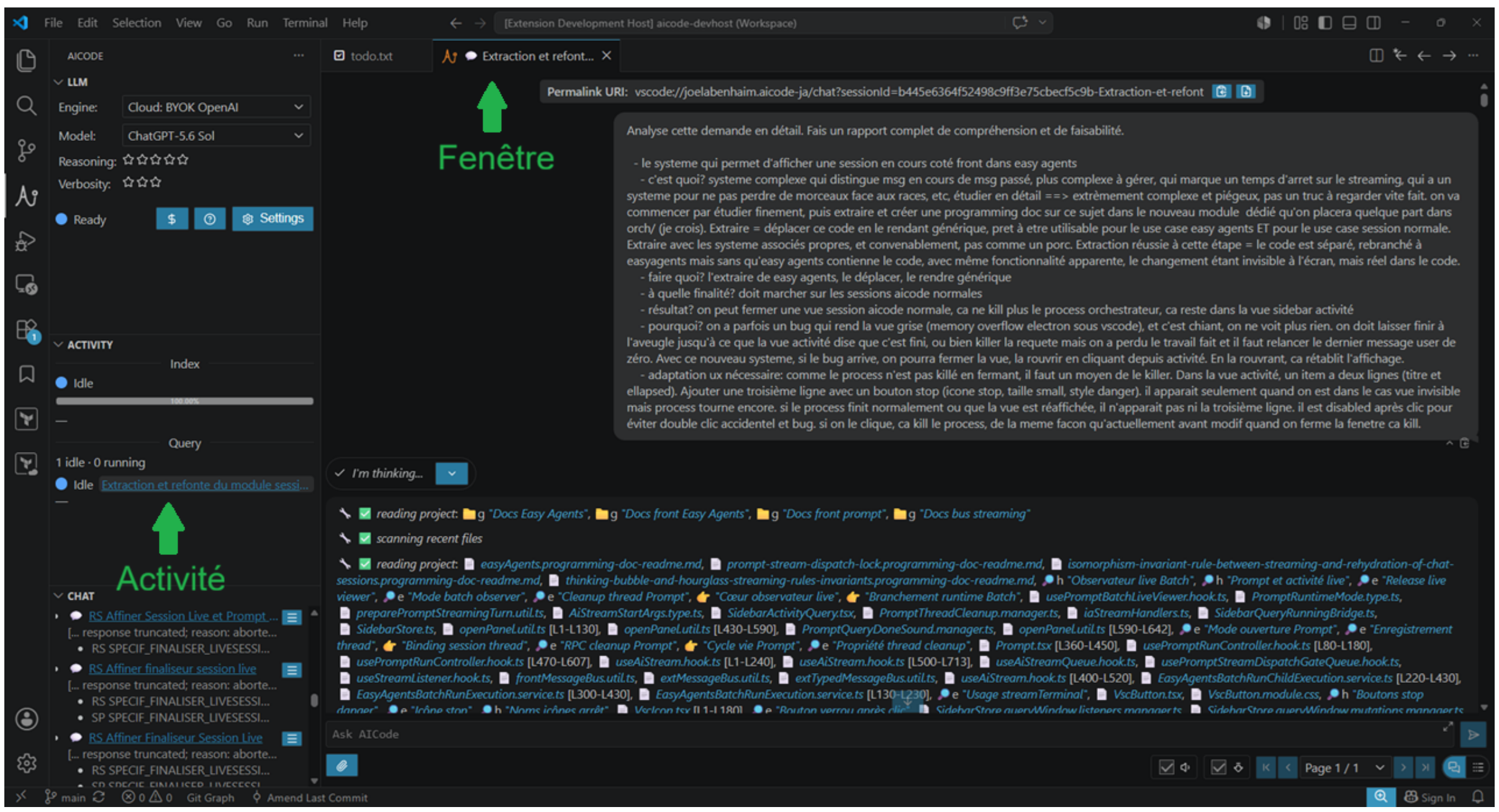

*Figure 1. The two interface regions involved: the activity sidebar (left) and the session panel that must become closable (centre).*

The requirement violates a structural guarantee on which the surrounding code was built: that a panel opens at the start of a request and remains open for its entire duration. Removing that guarantee affects request lifetime, memory ownership, cache restoration, ordering of asynchronous events, and the reattachment of a partially consumed stream.

The specific difficulties were: reconstructing a composite display, response text, reasoning traces, tool calls and code blocks interleaved in one view, from a partially consumed stream; race conditions between panel closure and in-flight generation; and the three-way contention, on reopening, between panel initialisation, replay of already-generated content, and live streaming of new tokens.

In the author's assessment, reconstructing a system's core end to end while dismantling a central invariant is the class of task that conventionally calls for rewriting the affected components from scratch, a more realistic path than attempting it as a refactor. The frozen specification produced for this task, itself the product of fourteen refinement cycles and published in full below, gives a direct, concrete view of that complexity. No formal complexity metric was computed; the specification itself is offered as the evidence.

Attached specification: [liveSession-logs-phase2-specification-en.pdf](liveSession-logs-phase2-specification-en.pdf) (55 pages).

The task was scoped in two slices. Slice 1 extracted the existing streaming logic from a subsystem where it was embedded, and generalised it into a reusable library, with no visible change in behaviour. Slice 2, the subject of this paper, applied that library to standard sessions and implemented the open/close behaviour.

The initial request was given to the agent in natural language, describing the user-visible problem, the intended outcome, and the interface adaptation required. It did not specify types, modules, protocols, or the asynchronous operations to be handled. Determining those was part of the work under study.

## 5. Method

The agent used is AICode, running the OpenAI ChatGPT 5.6 Sol model in max reasoning mode. The agent applies a five-phase workflow: ideate, specify, refine, code, verify. Each phase runs as a separate session. The volume and interdependence of the changes made a line-by-line review unrealistic given their complexity; the result depended on trust in the process rather than inspection of the code itself. Figure 2 and Table 1 summarise the five phases, what each produces, and what it is checked against.

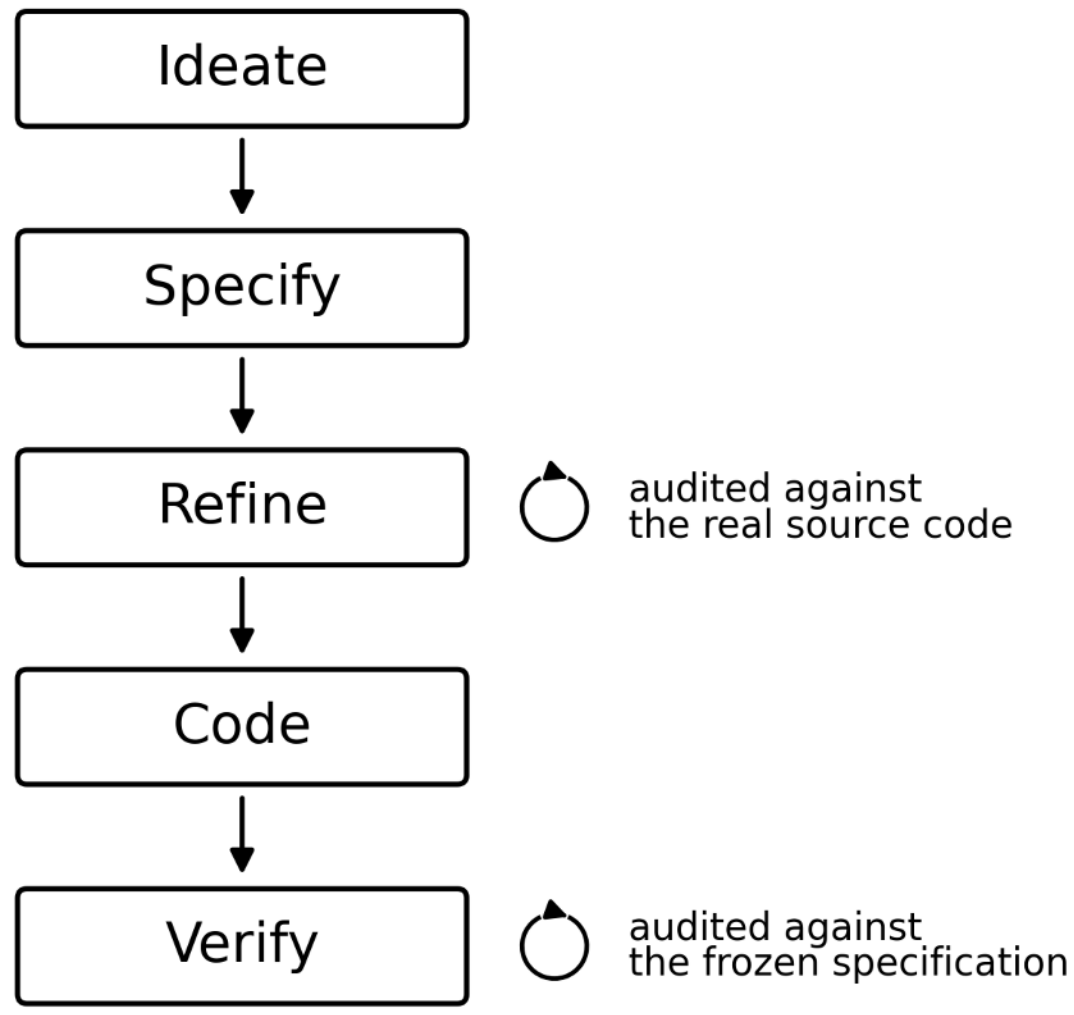

*Figure 2. The five-phase protocol applied in this study.*

Two of the five phases are loops that repeat until they stop finding anything to correct: refine re-audits the specification against the real source code; verify re-audits the generated code against that same specification, frozen once refinement ends.

*Table 1. The five phases, the artifact each produces, and what it is checked against.*

| Phase | Artifact produced | Checked against | Decision |
|---|---|---|---|
| Ideate | Verbalised intent and constraints | The real source code | Operator confirms scope |
| Specify | Formal specification | None (first production) | Operator skims through the specification |
| Refine (×14) | Revised specification | The real source code | Operator relaunches, or freezes |
| Code | Patches | The frozen specification | None (patches applied without inspection) |
| Verify (×17) | Corrections | The frozen specification | Operator accepts, or stops at two consecutive zero-finding passes |

## 5.1. Ideation and specification

From the initial request and one scoping exchange, the agent produced a formal specification describing changes to 110 files across shared contracts, backend, frontend and the RPC bus.

Attached session log: liveSession-logs-phase2-ideation.pdf (83 pages).

## 5.2. Refinement cycles

Refinement is the process of having the agent re-analyse the specification against the real source code, report defects in the specification, and rewrite it. Each cycle is a separate session, averaging 35 minutes.

Fourteen cycles were run, producing approximately 85 corrections to the specification and expanding its scope from 110 to 160 affected files as previously unnoticed dependencies were identified. Cycle 14 returned no findings, so the specification produced at cycle 13 was frozen and used as the reference for all subsequent phases.

Note on measurement: a platform-side migration issue prevented the correction list from being recovered for some cycles. In the table below, those cycles are recorded as an estimate of five.

*Table 2. Specification refinement cycles.*

| Cycle | Corrections | Files in scope | Log |
|---|---|---|---|
| 1 | ≈ 5 | 120 | liveSession-logs-phase2-refine-01.pdf (19 pages) |
| 2 | ≈ 5 | 122 | liveSession-logs-phase2-refine-02.pdf (24 pages) |
| 3 | 5 | 122 | liveSession-logs-phase2-refine-03.pdf (25 pages) |
| 4 | 4 | 122 | liveSession-logs-phase2-refine-04.pdf (19 pages) |
| 5 | 6 | 122 | liveSession-logs-phase2-refine-05.pdf (17 pages) |
| 6 | 10 | 129 | liveSession-logs-phase2-refine-06.pdf (21 pages) |
| 7 | ≈ 5 | 129 | liveSession-logs-phase2-refine-07.pdf (24 pages) |

| Cycle | Corrections | Files in scope | Log |
|---|---|---|---|
| 8 | ≈ 5 | 133 | liveSession-logs-phase2-refine-08.pdf (29 pages) |
| 9 | ≈ 5 | 149 | liveSession-logs-phase2-refine-09.pdf (32 pages) |
| 10 | 11 | 151 | liveSession-logs-phase2-refine-10.pdf (35 pages) |
| 11 | 7 | 154 | liveSession-logs-phase2-refine-11.pdf (38 pages) |
| 12 | 9 | 154 | liveSession-logs-phase2-refine-12.pdf (42 pages) |
| 13 | 8 | 160 | liveSession-logs-phase2-refine-13.pdf (25 pages) |
| 14 | 0 | 160 | liveSession-logs-phase2-refine-14.pdf (2 pages) |

*Total: approximately 85 specification corrections.*

Attached final specification #13: liveSession-logs-phase2-specification-en.pdf (55 pages).

### 5.3. Implementation

The agent was asked to implement the frozen specification. It declined the request in its initial form, stating that a partial implementation would leave the repository in an inconsistent state and violate the atomicity requirement of the specification:

> *"I did not apply any changes to the repository. [...] This migration affects 160 files, including 31 new ones, with interdependent changes across shared contracts, the backend registry, transport, persistence, panels, SidebarStore, and the React runtime. A partial implementation would have left the repository in an inconsistent state and violated the specification's atomicity requirement. Therefore, I did not introduce a half-migration that could have broken Prompt or EasyAgents."*

The agent proposed a ten-step decomposition; three steps were requested instead, and accepted. Implementation ran in three consecutive segments, each initiated by an explicit human confirmation, for a total of 2 hours 21 minutes 44 seconds of agent time.

Attached session log: liveSession-logs-phase2-code.pdf (323 pages).

### 5.4. Feedback loop

The agent does not automatically correct compilation, typing or unit-test failures in its own output; this is a deliberate design choice, intended to keep those signals visible. The residual syntax, typing and unit-test errors were corrected in a separate session in under one hour.

Attached session log: liveSession-logs-phase2-feedbackloop.pdf (61 pages).

### 5.5. Verification cycles

Verification is the process of having the agent analyse the current state of the source code against the frozen specification, report residual architectural defects and subtle deviations, and correct them. It differs from the feedback loop in that it targets architectural conformity rather than compilation validity.

Seventeen cycles were run, producing 116 code corrections. Each cycle took between 7 and 45 minutes for the audit itself, plus 15 to 60 minutes for automatic generation of the corrections.

*Table 3. Code verification cycles.*

| Cycle | Corrections | Log |
|---|---|---|
| 1 | 10 | liveSession-logs-phase2-verify-01.pdf (53 pages) |
| 2 | 12 | liveSession-logs-phase2-verify-02.pdf (102 pages) |
| 3 | 9 | liveSession-logs-phase2-verify-03.pdf (102 pages) |
| 4 | 21 | liveSession-logs-phase2-verify-04.pdf (152 pages) |
| 5 | 2 | liveSession-logs-phase2-verify-05.pdf (9 pages) |
| 6 | 7 | liveSession-logs-phase2-verify-06.pdf (64 pages) |
| 7 | 8 | liveSession-logs-phase2-verify-07.pdf (61 pages) |
| 8 | 13 | liveSession-logs-phase2-verify-08.pdf (34 pages) |
| 9 | 6 | liveSession-logs-phase2-verify-09.pdf (51 pages) |
| 10 | 3 | liveSession-logs-phase2-verify-10.pdf (33 pages) |
| 11 | 3 | liveSession-logs-phase2-verify-11.pdf (29 pages) |
| 12 | 4 | liveSession-logs-phase2-verify-12.pdf (34 pages) |
| 13 | 4 | liveSession-logs-phase2-verify-13.pdf (35 pages) |
| 14 | 4 | liveSession-logs-phase2-verify-14.pdf (43 pages) |
| 15 | 10 | liveSession-logs-phase2-verify-15.pdf (25 pages) |
| 16 | 0 | liveSession-logs-phase2-verify-16.pdf (6 pages) |
| 17 | 0 | liveSession-logs-phase2-verify-17.pdf (21 pages) |

*Total: 116 code corrections.*

### 5.6. Convergence criterion

The stopping rule was empirical rather than fixed in advance: verification continued until two consecutive cycles returned zero findings. That condition was met at cycles 16 and 17.

Phase 2 was executed end to end without the program being run once. The first manual execution took place after the seventeenth verification cycle.

## 6. Results

### 6.1. Defects removed before execution

Across the 31 audit passes, 201 defects, ambiguities and architectural deviations were identified and corrected before any human ran the code: approximately 85 in the specification, 116 in the generated code.

### 6.2. Result

The operation touched 189 files in three days: more than the 160 tracked by the frozen specification (§5.2), whose count reflects planned scope at freeze time rather than the additional files touched during implementation and the seventeen verification cycles that followed.

After the operation, the codebase comprised ≈ 736,000 lines. The two commits corresponding to the extraction and to the open/close refactor are as follows.

```
commit ed3bd94687cc34f6603730689bfc775818881bae
refactor(live-session): extract EasyAgents live watcher into generic orch modules
99 files changed, 11107 insertions(+), 4572 deletions(-)

commit 23c18d4f4e8757c06cc633bd9e014989b21285a8
feat(live-session): keep prompt runs alive across panel close and reopen
189 files changed, 23663 insertions(+), 11850 deletions(-)
```

*Listing 1. Git commit log for the two commits comprising this operation: the Slice 1 extraction and the Slice 2 open/close refactor.*

Combined: 288 changed files, 34,770 insertions, 16,422 deletions.

## 6.3. Functional validation

On first manual execution, the specified behaviour was present and no defect or regression was observed: panel closing and reopening during streaming, display restoration, the detached stop control, and the pre-existing subsystem behaviour all operated as specified. In addition to this manual check, the project's large pre-existing unit test suite, predating the refactor and exercising behaviour across both affected and unaffected modules, was run after the change and showed no regression. The software has since been used across roughly thirty further sessions covering its different use cases, with no bug observed in any of them. Automated test coverage was maintained throughout, tests being written by the agent under a standing instruction that all produced code must be tested.

The build is publicly downloadable, and the streaming-resumption behaviour it reports, closing and reopening a panel without interrupting or duplicating the underlying generation, can be verified directly by any user rather than taken on the author's account alone.

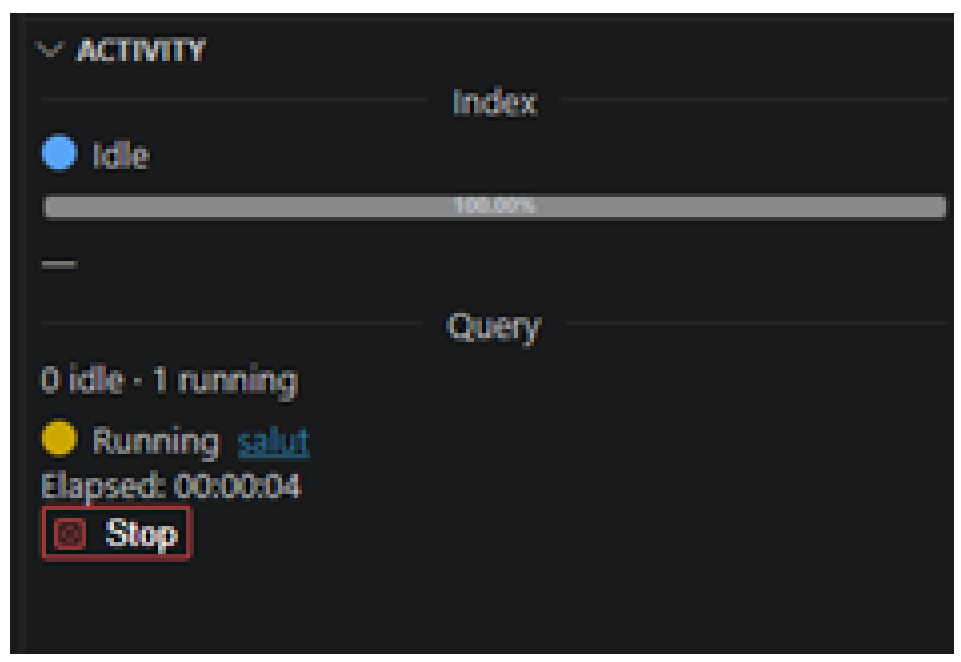

*Figure 3. Activity sidebar after the refactor: a session whose panel has been closed while its generation continues, showing elapsed time and the detached stop control.*

The build was released the same day as version 2.3.0 and is publicly available, allowing independent functional inspection of the outcome, though not of the process.

## 6.4. Subsequent adjustment

One retrospective interface decision was revised after release. The change required modifying a single source file plus its tests. A change confined to one file is generally taken as an indication that responsibility for that behaviour was not scattered across the codebase, and could be considered as an indication of architectural cleanliness.

Attached session log: liveSession-logs-phase2-verify-17.pdf (21 pages).

### 6.5. Cost

The three-day operation consumed USD 2,430 in model inference.

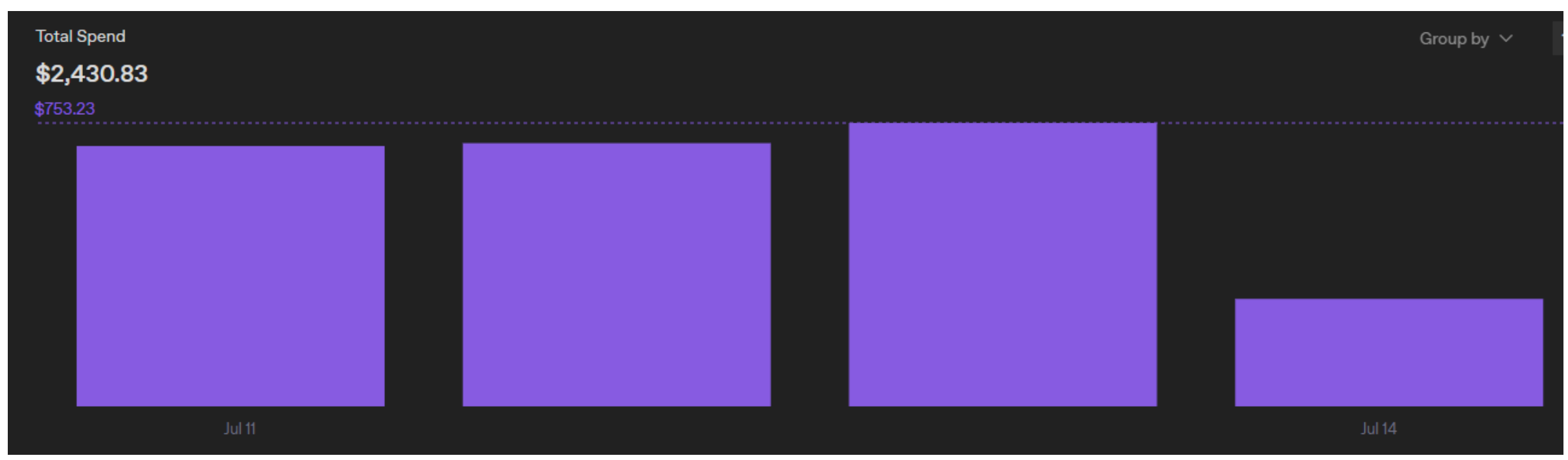

*Figure 4. Daily inference spend across the operation, totalling USD 2,430.*

## 7. Discussion

A written specification artifact, challenged fourteen times against the source code, executed literally, then audited seventeen times against itself, did converge towards conformity by repetition. This does not require the model to be individually reliable on any single pass.

Defects are cheaper to remove earlier: a defect caught in the specification costs a paragraph to fix, the same defect caught after generation costs a set of interdependent code changes. This is what the refinement cycles are for.

Verification works differently: the code is checked, session after session, against the frozen specification, a fixed target, until no more deviations are found.

The frozen specification itself, product of the fourteen refinement cycles described in Section 5.2, is the artifact this argument rests on; it is published in full rather than summarised.

Attached specification: liveSession-logs-phase2-specification-en.pdf (55 pages).

## 8. Scope of the claim

To situate the class of task precisely, this case study is not:

- A project written from scratch.
- A language-to-language transpilation, where a source implementation acts as an oracle.
- A framework or version migration following a documented upgrade path.
- A reimplementation of an open-source project likely present in the model's training data.

It is a maintenance operation on an original problem for which, to the author's knowledge, no directly comparable open-source case exists, in which the target behaviour had to be derived rather than transposed, and in which no pre-existing test suite could serve as an oracle for the new behaviour, that behaviour not having existed before.

## 9. Limitations

The limitations of this report are structural and are stated in full.

1. Single case. One task, one codebase, one operator. Nothing here establishes a distribution of outcomes, a success rate, or reproducibility across other systems or other classes of problem.
2. No control condition. The same task was not attempted with other agents under comparable conditions. Doing so credibly would require operators expert in those tools, since piloting a competing agent without that expertise would bias the comparison in either direction.
3. Self-reported. The author designed the tool, performed the operation, and reports the result. The raw logs are published to allow inspection, but the reporting itself is not independent.
4. Definition of "no bug observed". The claim refers to observed behaviour on first manual execution, across roughly thirty subsequent sessions of use, and the automated test suite. It is not a proof of absence of latent defects, which no test campaign or usage window of any length establishes.
5. Closed source. The codebase cannot be published, so the operation cannot be replayed by a third party on the same material.
6. Model dependency. The result was obtained with a specific frontier model in extended reasoning mode. The protocol's behaviour with weaker models is not characterised here.
7. Instrumented scope. The published logs cover the reported operation only. The claim that the whole system was built under the same methodology is the author's declaration: a 4,000-commit history cannot be instrumented retroactively.

The clearest next step would be execution of the same protocol by an independent operator on a public codebase, which would remove the first, third and fifth limitations at once.

## 10. Data availability

The complete raw session logs of every cycle described in this paper are published, together with the frozen specification. They total 1,500+ pages and are in French. They can be read directly, or submitted to a language model for consistency checking against the claims made here.

Every attached document referenced in this paper is listed below, in the order it is first cited.

| Attached files |
|---|
| https://aisovereignlabs.ai/docs/case-study/liveSession/logs/ (Data availability index page) |
| https://aisovereignlabs.ai/docs/case-study/liveSession/logs/liveSession-logs-phase2-specification-en.pdf (55 pages) |
| https://aisovereignlabs.ai/docs/case-study/liveSession/logs/liveSession-logs-phase2-ideation.pdf (83 pages) |
| https://aisovereignlabs.ai/docs/case-study/liveSession/logs/liveSession-logs-phase2-refine-01.pdf (19 pages) |
| https://aisovereignlabs.ai/docs/case-study/liveSession/logs/liveSession-logs-phase2-refine-02.pdf (24 pages) |
| https://aisovereignlabs.ai/docs/case-study/liveSession/logs/liveSession-logs-phase2-refine-03.pdf (25 pages) |
| https://aisovereignlabs.ai/docs/case-study/liveSession/logs/liveSession-logs-phase2-refine-04.pdf (19 pages) |
| https://aisovereignlabs.ai/docs/case-study/liveSession/logs/liveSession-logs-phase2-refine-05.pdf (17 pages) |
| https://aisovereignlabs.ai/docs/case-study/liveSession/logs/liveSession-logs-phase2-refine-06.pdf (21 pages) |

| Attached files |
| --- |
| https://aisovereignlabs.ai/docs/case-study/liveSession/logs/liveSession-logs-phase2-refine-07.pdf (24 pages) |
| https://aisovereignlabs.ai/docs/case-study/liveSession/logs/liveSession-logs-phase2-refine-08.pdf (29 pages) |
| https://aisovereignlabs.ai/docs/case-study/liveSession/logs/liveSession-logs-phase2-refine-09.pdf (32 pages) |
| https://aisovereignlabs.ai/docs/case-study/liveSession/logs/liveSession-logs-phase2-refine-10.pdf (35 pages) |
| https://aisovereignlabs.ai/docs/case-study/liveSession/logs/liveSession-logs-phase2-refine-11.pdf (38 pages) |
| https://aisovereignlabs.ai/docs/case-study/liveSession/logs/liveSession-logs-phase2-refine-12.pdf (42 pages) |
| https://aisovereignlabs.ai/docs/case-study/liveSession/logs/liveSession-logs-phase2-refine-13.pdf (25 pages) |
| https://aisovereignlabs.ai/docs/case-study/liveSession/logs/liveSession-logs-phase2-refine-14.pdf (2 pages) |
| https://aisovereignlabs.ai/docs/case-study/liveSession/logs/liveSession-logs-phase2-code.pdf (323 pages) |
| https://aisovereignlabs.ai/docs/case-study/liveSession/logs/liveSession-logs-phase2-feedbackloop.pdf (61 pages) |
| https://aisovereignlabs.ai/docs/case-study/liveSession/logs/liveSession-logs-phase2-verify-01.pdf (53 pages) |
| https://aisovereignlabs.ai/docs/case-study/liveSession/logs/liveSession-logs-phase2-verify-02.pdf (102 pages) |
| https://aisovereignlabs.ai/docs/case-study/liveSession/logs/liveSession-logs-phase2-verify-03.pdf (102 pages) |
| https://aisovereignlabs.ai/docs/case-study/liveSession/logs/liveSession-logs-phase2-verify-04.pdf (152 pages) |
| https://aisovereignlabs.ai/docs/case-study/liveSession/logs/liveSession-logs-phase2-verify-05.pdf (9 pages) |
| https://aisovereignlabs.ai/docs/case-study/liveSession/logs/liveSession-logs-phase2-verify-06.pdf (64 pages) |
| https://aisovereignlabs.ai/docs/case-study/liveSession/logs/liveSession-logs-phase2-verify-07.pdf (61 pages) |
| https://aisovereignlabs.ai/docs/case-study/liveSession/logs/liveSession-logs-phase2-verify-08.pdf (34 pages) |
| https://aisovereignlabs.ai/docs/case-study/liveSession/logs/liveSession-logs-phase2-verify-09.pdf (51 pages) |
| https://aisovereignlabs.ai/docs/case-study/liveSession/logs/liveSession-logs-phase2-verify-10.pdf (33 pages) |
| https://aisovereignlabs.ai/docs/case-study/liveSession/logs/liveSession-logs-phase2-verify-11.pdf (29 pages) |
| https://aisovereignlabs.ai/docs/case-study/liveSession/logs/liveSession-logs-phase2-verify-12.pdf (34 pages) |
| https://aisovereignlabs.ai/docs/case-study/liveSession/logs/liveSession-logs-phase2-verify-13.pdf (35 pages) |
| https://aisovereignlabs.ai/docs/case-study/liveSession/logs/liveSession-logs-phase2-verify-14.pdf (43 pages) |
| https://aisovereignlabs.ai/docs/case-study/liveSession/logs/liveSession-logs-phase2-verify-15.pdf (25 pages) |
| https://aisovereignlabs.ai/docs/case-study/liveSession/logs/liveSession-logs-phase2-verify-16.pdf (6 pages) |
| https://aisovereignlabs.ai/docs/case-study/liveSession/logs/liveSession-logs-phase2-verify-17.pdf (21 pages) |

The software described in this paper is publicly downloadable [14] in two versions, allowing direct verification of the described behaviour change. In any version below 2.3.0, starting a query and then closing its panel terminates the query. In version 2.3.0 and above, the same action preserves the query, and clicking the session name reopens the live panel, reattached to the same ongoing stream.

## Competing interests

The author's company, AI Sovereign Labs, designed and distributes the agent used in this study. The published logs are provided so the process can be inspected directly.